\documentclass[a4paper,11pt]{article}
\usepackage{pos}
\usepackage[english]{babel}       %Dictionary for English special characters
\usepackage{color}                %Color characters
\usepackage{titlesec}             %Select alternative section titles
\usepackage{newlfont}          	  %Font commands
\usepackage[T1]{fontenc}          %Package for accents

\usepackage{tikz}              %Package to draw plots and diagrams
\usepackage{tikz-feynman}      %Package to draw Feynman diagrams
\tikzfeynmanset{compat=1.1.0}  %Setting for Feynman diagrams
\usetikzlibrary{bending}       %Library to bend arrows in tikz
\usetikzlibrary{decorations.markings} %Library to include decorations for lines in tikz
\usetikzlibrary {arrows.meta} %Library to include Arrows for lines in Tikz
\usetikzlibrary{snakes}        %Library to include waives lines in tikz
\usepackage{graphicx}          %Package to include figures
\usepackage{pdfpages}          %Package to include pdfs

\usepackage{amsmath}  %Mathematical formule
\usepackage{braket}   %Dirac Brackets
\usepackage{cancel}   %Cancel terms in formula
\usepackage{dsfont}   %Mathematical symbols
\usepackage{array}    %Package for arrays
\usepackage{multicol} %Package for multicolumn cells
\usepackage{makecell} %Package to break lines in cells

\pgfdeclaredecoration{Snake}{initial}
{\state{initial}[switch if less than=+.625\pgfdecorationsegmentlength to final,
                  width=+.3125\pgfdecorationsegmentlength,
                  next state=up]{
    \pgfpathmoveto{\pgfqpoint{0pt}{\pgfdecorationsegmentamplitude}}
  }
  \state{down}[switch if less than=+.8125\pgfdecorationsegmentlength to end down,
               width=+.5\pgfdecorationsegmentlength,
               next state=up]{
    \pgfpathcosine{\pgfqpoint{.25\pgfdecorationsegmentlength}{-1\pgfdecorationsegmentamplitude}}
    \pgfpathsine{\pgfqpoint{.25\pgfdecorationsegmentlength}{-1\pgfdecorationsegmentamplitude}}
  }
  \state{up}[switch if less than=+.8125\pgfdecorationsegmentlength to end up,
             width=+.5\pgfdecorationsegmentlength,
             next state=down]{
    \pgfpathcosine{\pgfqpoint{.25\pgfdecorationsegmentlength}{\pgfdecorationsegmentamplitude}}
    \pgfpathsine{\pgfqpoint{.25\pgfdecorationsegmentlength}{\pgfdecorationsegmentamplitude}}
  }
  \state{end down}[width=+.3125\pgfdecorationsegmentlength,
                   next state=final]{
     \pgfpathcosine{\pgfqpoint{.25\pgfdecorationsegmentlength}{-1\pgfdecorationsegmentamplitude}}
     \pgfpathsine{\pgfqpoint{.25\pgfdecorationsegmentlength}{-1\pgfdecorationsegmentamplitude}}
  }
  \state{end up}[width=+.3125\pgfdecorationsegmentlength,
                 next state=final]{
     \pgfpathcosine{\pgfqpoint{.25\pgfdecorationsegmentlength}{1\pgfdecorationsegmentamplitude}}
     \pgfpathsine{\pgfqpoint{.25\pgfdecorationsegmentlength}{1\pgfdecorationsegmentamplitude}}
  }
  \state{final}{\pgfpointdecoratedpathlast}
}

\title{Hutch++ and XTrace to Improve Stochastic Trace Estimation}
\author*[a]{Alessandro Cotellucci}
\author[a,b]{Agostino Patella}

\affiliation[a]{Humboldt Universit\"at zu Berlin, Institut f\"ur Physik \& IRIS Adlershof, \\Zum Grossen Windkanal 6, 12489 Berlin, Germany}
\affiliation[b]{DESY, Platanenallee 6, D-15738 Zeuthen, Germany}

\emailAdd{alecote@physik.hu-berlin.de}
\abstract{We present the analysis of two recently proposed noise reduction techniques, Hutch++ and XTrace, both based on inexact deflation. These methods were proven to have a better asymptotic convergence to the solution than the classical Girard-Hutchinson stochastic method. We applied these methods to the computation of the trace of the inverse of the Dirac operator with $O(a)$ improved Wilson fermions on the QCD ensemble generated by the RC$^{\star}$ collaboration with $m_{\pi}\approx 400$ MeV and $V = 64\times 32^3$. Unfortunately, we see no noise reduction with a moderate number of sources, and we attempt an explanation of why this is the case.
This study was part of the effort to evaluate isospin-breaking effects using the RM123 with C$^{\star}$ boundary conditions in an unquenched set-up.}

\FullConference{The 40th International Symposium on Lattice Field Theory (Lattice 2023)\\
July 31st - August 4th, 2023\\
Fermi National Accelerator Laboratory\\}

\begin{document}
\maketitle

\section{Motivations}
The evaluation of quark disconnected diagrams is a crucial problem in many domains of Lattice QCD. In the context of isospin-breaking effects with the RM123 method \cite{deDivitiis:2013xla} a collection of these diagrams arise from the sea-sea contractions. The mentioned diagrams have the following topology: 
\begin{equation}
\begin{aligned}
&\text{Strong IB:}\ \text{Tr}\left[D^{-1}\right]=\vcenter{\hbox{\begin{tikzpicture}[>={Triangle[bend,width=2pt,length=3pt]}]
        \coordinate (z);
               \node at (z)[circle,fill,inner sep=0.6pt]{};
        \draw (z) arc[start angle=0,end angle=360,radius=0.25];
        \draw[->] (z) arc[start angle=0,end angle=200,radius=0.25];
\end{tikzpicture}}}\\
&\text{Tadpole:}\ \text{Tr}\left[TD^{-1}\right]=\vcenter{\hbox{\begin{tikzpicture}[>={Triangle[bend,width=2pt,length=3pt]}]
        \coordinate (z);
        \coordinate[below left = 0.0094 and 0.033 of z] (w);
        \node at (z)[circle,fill,inner sep=0.6pt]{};
        \draw[decoration=Snake,segment length=2.9342574999999pt,segment amplitude=0.9pt, decorate] (w) arc[start angle=-180,end angle=180,radius=0.14999878];        
        \draw (z) arc[start angle=0,end angle=360,radius=0.25];
        \draw[->] (z) arc[start angle=0,end angle=200,radius=0.25];
\end{tikzpicture}}}\\
&\text{Bubbles:}\ \text{Tr}\left[JD^{-1}\right]\text{Tr}\left[JD^{-1}\right]=\vcenter{\hbox{\begin{tikzpicture}[>={Triangle[bend,width=2pt,length=3pt]}]
        \coordinate (z);
        \coordinate[right=0.75 of z] (w);
        \node at (w)[circle,fill,inner sep=0.6pt]{};
        \node at (z)[circle,fill,inner sep=0.6pt]{};
        \draw (z) arc[start angle=0,end angle=360,radius=0.25]
         (w) arc[start angle=-180,end angle=180,radius=0.25];
        \draw[->] (z) arc[start angle=0,end angle=200,radius=0.25];
        \draw[->] (w) arc[start angle=180,end angle=380,radius=0.25];
        \draw[decoration=snake,segment length=2.9pt,segment amplitude=0.9pt,decorate] (z) -- (w);
\end{tikzpicture}}}\\
&\text{Self-Energy:}\ \text{Tr}\left[JD^{-1}JD^{-1}\right]=\vcenter{\hbox{\begin{tikzpicture}[>={Triangle[bend,width=2pt,length=3pt]}]
        \coordinate (z);
        \coordinate[below=0.5 of z] (w);
        \node at (w)[circle,fill,inner sep=0.6pt]{};
        \node at (z)[circle,fill,inner sep=0.6pt]{};      
        \draw (z) arc[start angle=90,end angle=450,radius=0.25];
        \draw[->] (z) arc[start angle=90,end angle=200,radius=0.25];
        \draw[->] (w) arc[start angle=270,end angle=380,radius=0.25];
        \draw[decoration=snake,segment length=2.9pt,segment amplitude=0.9pt,decorate] (z) -- (w);  
\end{tikzpicture}}}
\end{aligned}\label{eq:diagrams}
\end{equation}
where $D^{-1}$ is the inverse of the Dirac operator, $J$ is the conserved electromagnetic current and $T$ is the Tadpole operator which arises from the lattice discretization of QED. The trace in \eqref{eq:diagrams}  runs over the spin, colour and spacetime indices of the matrices and due to a large number of entries it can not be evaluated exactly but has to be approximated stochastically.\\
The straightforward way to compute this quantity is by using the Girard-Hutchinson trace estimator \cite{Girard1989, doi:10.1080/03610919008812866}. Recently, two new methods have been proposed:  the Hutch++ \cite{doi:10.1137/1.9781611976496.16} and the XTrace \cite{epperly2023xtrace} which modify the Girard-Hutchinson trace estimator to get a better asymptotic scaling. The Hutch++ has already been studied in Lattice QCD simulations by \cite{Frommer:2022qiy} in the context of the MGMLMC++ trace estimator while in this study we focus on the Hutch++ as a substitute candidate for the Girard-Hutchinson. This is the first study involving the XTrace in Lattice QCD simulations. \\
Since the inversion of the Dirac operator is the most expensive operation in these algorithms, we use the number of inversions as the cost unit to compare the different methods and neglect the impact of all the other operations.
   
\section{Hutch++}
The stochastic estimator used to evaluate the trace with this method is:
\begin{equation}
\langle\text{Tr}\left[D^{-1}\right]\rangle =\underbrace{\sum_{i}^{n_{vec}}q_i^{\dagger}D^{-1}q_i}_{\text{Deflated term}}
+\underbrace{\frac{1}{n_{src}}\sum_i^{n_{src}} g_i^{\dagger}\left(\mathds{1}-\sum_jq_jq_j^{\dagger}\right)D^{-1}g_i}_{\text{Stochastic remnant}}
\end{equation}
where $D^{-1}$ is the inverse of the Dirac operator, $g_i$ are U(1) random sources and $q_i$ are the vectors belonging to the deflation space. The deflation space is obtained by the orthonormalization of the vectors $D^{-1}w_i$, where $w_i$ U(1) are random sources.\\
The number of inversions of the Dirac operator is then fixed by $m = 2 n_{vec} + n_{src}$.
The effect of the deflation space is to give a rough approximation of the lowest eigenmodes of $D$ in the fashion of the low-modes averaging, even with this rough eigenspace the algorithm has been proved to always have a better convergence in the asymptotic regime than the Girard-Hutchinson estimator.

Since we are interested in the real part of $\text{Tr}\left[D^{-1}\right]$ we test also a modification of the method where the deflation space is computed with the hermitian part of the inverse of the Dirac operator $D^{-1}+D^{-1\dagger}$ for which the number of inversions is $m = 3 n_{vec} + n_{src}$.\\
In \cite{doi:10.1137/1.9781611976496.16} the authors propose an optimal choice for $n_{vec}$ and $n_{src}$ which minimizes the variance, in our case the optimal parameters are defined by:
\begin{equation}
\begin{aligned}
&\text{\makecell{Operator used to \\ compute deflation space}} && \text{Optimized parameters} && \text{Non optimized parameters} \\
& D^{-1} && n_{vec}=\frac{m+2}{4},\ n_{src}=\frac{m-2}{2}  && n_{vec}=n_{src}=\frac{m}{3} \\ 
&\left(D^{-1}+D^{-1\dagger}\right) && n_{vec}=\frac{m+2}{5},\ n_{src}=\frac{2m-6}{5} && n_{vec}=n_{src}=\frac{m}{4}.
\end{aligned}
\end{equation}

The two methods are tested to evaluate the trace of the inverse of the Dirac operator on an ensemble with $N_f=3+1$ $O\left(a\right)$ improved Wilson fermions with C$^{\star}$ boundary conditions in space and periodic boundary conditions in time \cite{RCstar:2022yjz}. 

\begin{table}[!ht]
   \centering
      \scalebox{0.8}{\begin{tabular}{cccccc}
         \hline
         lattice & $a$ [fm] & $m_{\pi}$ [MeV] & $m_{D}$ [MeV] & no. cnfg\\
         \hline
         \hline
         $64 \times 32^3$ & $0.05393(24)$  & $398.5(4.7)$ & $1912.7(5.7)$ & $1$\\
         \hline
      \end{tabular}}
  \caption{\label{tab:ens}Configuration used for this work from \cite{RCstar:2022yjz}.}
\end{table}

The ensemble characteristics are listed in Table \ref{tab:ens}, to focus only on the error of the estimator we decide to use only one gauge configuration.

\begin{figure}[!ht]
\begin{center}
  \includegraphics[width=11cm]{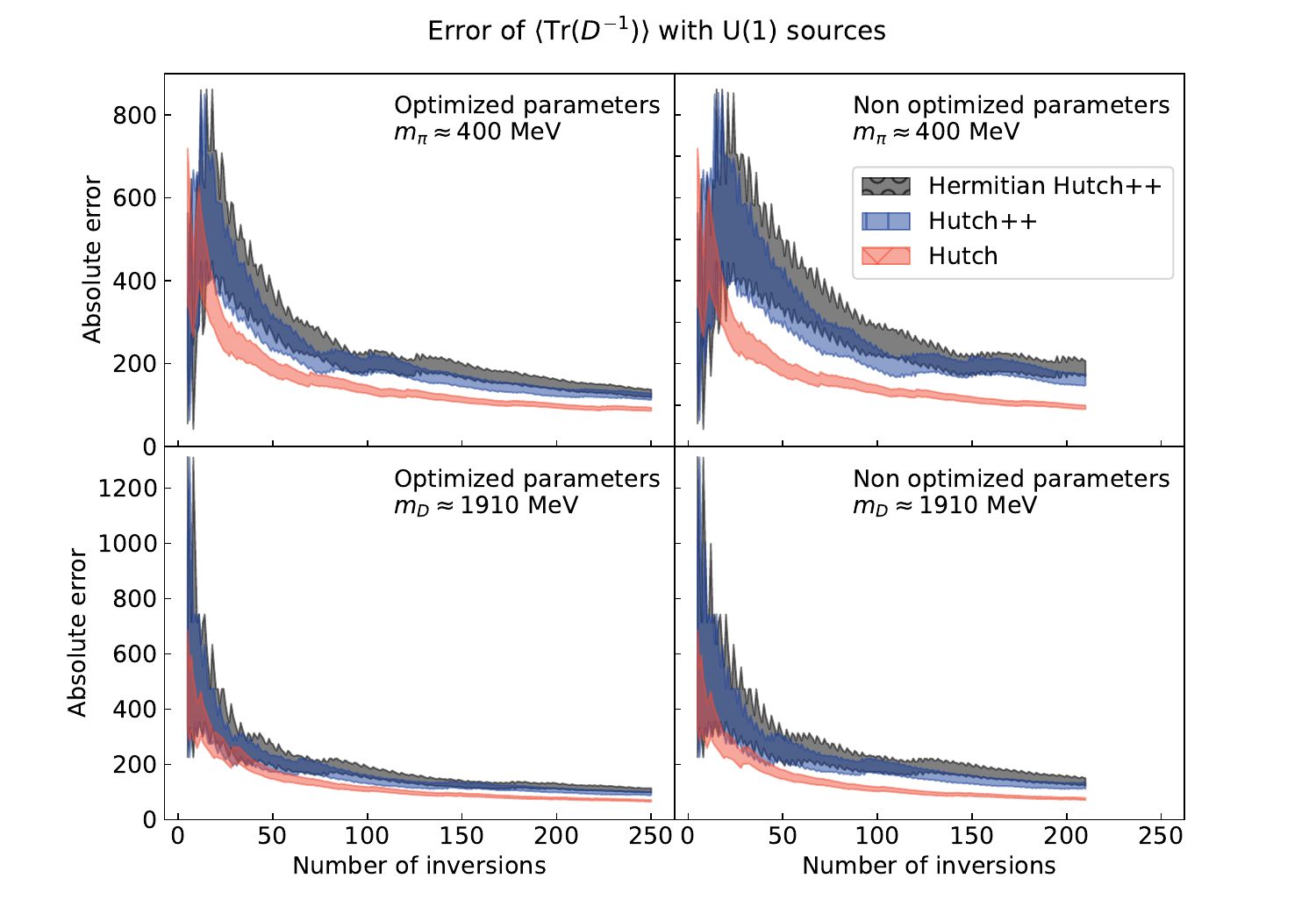}
\caption{\label{fig:HPP_Plot}Scaling of the absolute error as a function of the number of inversions of $D$ for the different algorithms using U(1) random sources for both the light quarks (top) and the charm quark (bottom).}
\end{center}
\end{figure}

In Figure \ref{fig:HPP_Plot} we show the results of our comparison for a maximum number of inversions of the Dirac operator of $O(200)$.\\
As one can see the scaling using the optimal parameters is confirmed to be better than the non optimized ones but in all the cases the Hutch++ fails to do better than the Girard-Hutchinson showing a similar scaling.

\subsection{The impact of the deflation space}
We study the impact of the deflation space by comparing the relative contribution of the deflated term to the total trace and the statistical error of the remnant as defined:
\begin{equation}
\begin{aligned}
&\text{Relative Deflated term:}\ &&\frac{\sum_{i}^{n_{vec}}q_i^{\dagger}D^{-1}q_i}{\langle\text{Tr}\left(D^{-1}\right)\rangle}\\
&\text{Relative Error:}\ &&\frac{\sigma\left[\text{Tr}\left(PD^{-1}\right) \right]}{\langle\text{Tr}\left(D^{-1}\right)\rangle},
\end{aligned}
\end{equation}
where $P$ is just the projector $\left(\mathds{1}-\sum_jq_jq_j^{\dagger}\right)$. In this case, we keep fixed the number of sources used to compute the stochastic remnant and we vary the number of inversions necessary to compute the Deflated term ($2n_{vec}$ for the Hutch++ and $3n_{vec}$ for the Hermitian Hutch++). When the relative deflated term is smaller than the relative error the actual contribution can not be distinguished from a statistical fluctuation.

\begin{figure}[!ht]
\begin{center}
 \includegraphics[width=14cm]{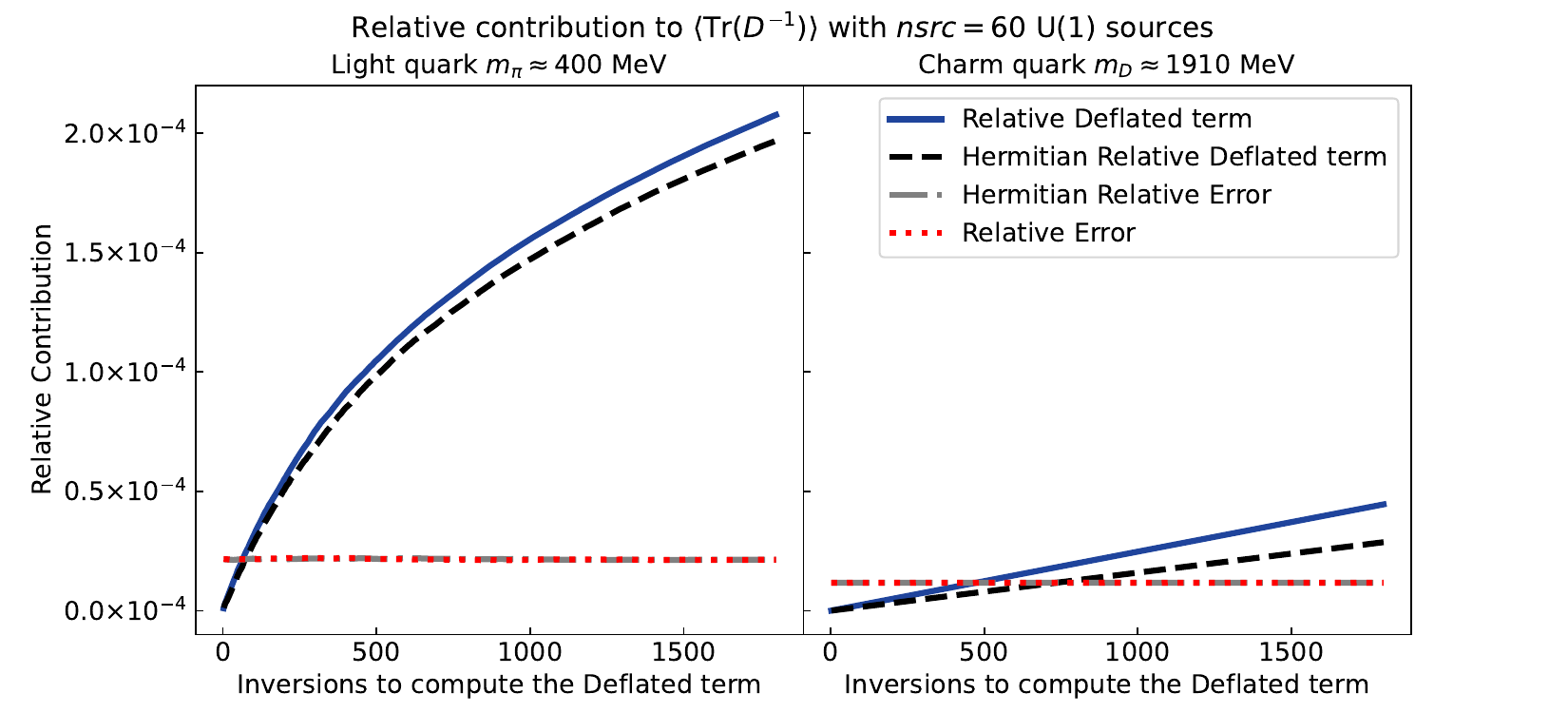}
	\caption{\label{fig:Nvec_plot}Scaling of the relative contribution of the different Hutch++ methods with the number of inversions to compute the deflation space. The blue solid line represents the Hutch++, the black dashed line the hermitian Hutch++, the red pointed line the relative statistical error in the Hutch++ case and the dashed green line the relative statistical error in the hermitian Hutch++ case.}
\end{center}
\end{figure} 

In Figure \ref{fig:Nvec_plot} one can see that the impact of the deflation space becomes relevant only after a relatively big amount of inversions $O(60)$ for the light quark and over $O(500)$ for the charm quark. There is a clear improvement in the effect going to smaller quark mass but the relative contribution of the deflation space is still of the order $O\left(10^{-4}\right)$ for an extremely high computational effort.

\begin{figure}[!ht]
\begin{center}
  \includegraphics[width=14cm]{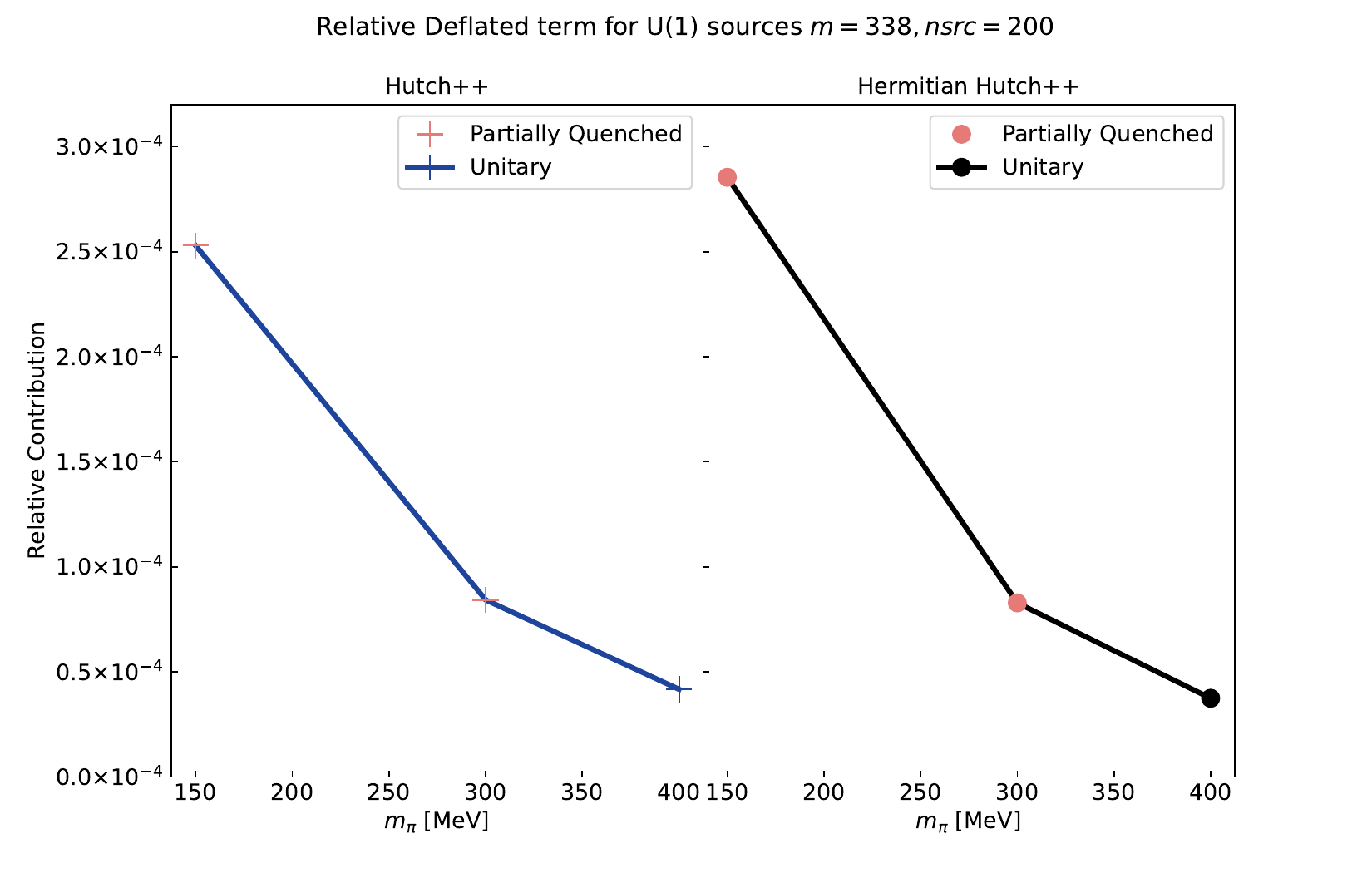}
\caption{\label{fig:Masses_Plot} Plot of the Relative contribution of the deflation term as a function of the pion mass for the Hutch++ (left) and the hermitian Hutch++ (right). The red points are the partially quenched setup while others are the unitary points.}
\end{center}
\end{figure}

Since the impact of the deflation space is bigger for the light quark than the charm one can expect the method to become more effective close to the physical pion mass like other eigenspace methods, to test this hypothesis we use a partially quenched setup. In Figure \ref{fig:Masses_Plot} we have the results with the relative contribution of the deflation term as a function of the pion mass. One can see that the behaviour of the Hutch++ and the hermitian Hutch++ is similar with a fixed number of inversions $m$ but the effect of the relative term is still irrelevant even at the physical point.

\newpage
\section{XTrace}
The other method which was recently proposed, and we decided to test, is the XTrace estimator. In this case, the form of the estimator is:
\begin{equation}
\langle\text{Tr}\left[D^{-1}\right]\rangle =\frac{1}{n_{src}}\sum_j^{n_{src}}\left[ \sum_{i}q_i^{\ j\dagger}D^{-1}q^{\ j}_i + w_j^{\dagger}\left(\mathds{1}-\sum_kq^{\ j}_kq_k^{\ j\dagger}\right)D^{-1}w_j\right]\notag
\end{equation}
where $w_i$ are U(1) random sources and the deflation vectors $q_i^{\ j}$ are obtained by the orthonormalization of $D^{-1}w_i$ with $i\neq j$ so all the sources are reused to construct the deflation space and the total number of inversions of $D$ is fixed to $m=2n_{src}$. Also, in this case, we add an implementation of the method where the subspace is generated using the hermitian part of $D^{-1}$ and in that case, the number of inversions is $m=3n_{src}$. As a modification of the Hutch++, these methods inherit better asymptotic scaling than the Girard-Hutchinson.

\begin{figure}[!ht]
\begin{center}
  \includegraphics[width=13cm]{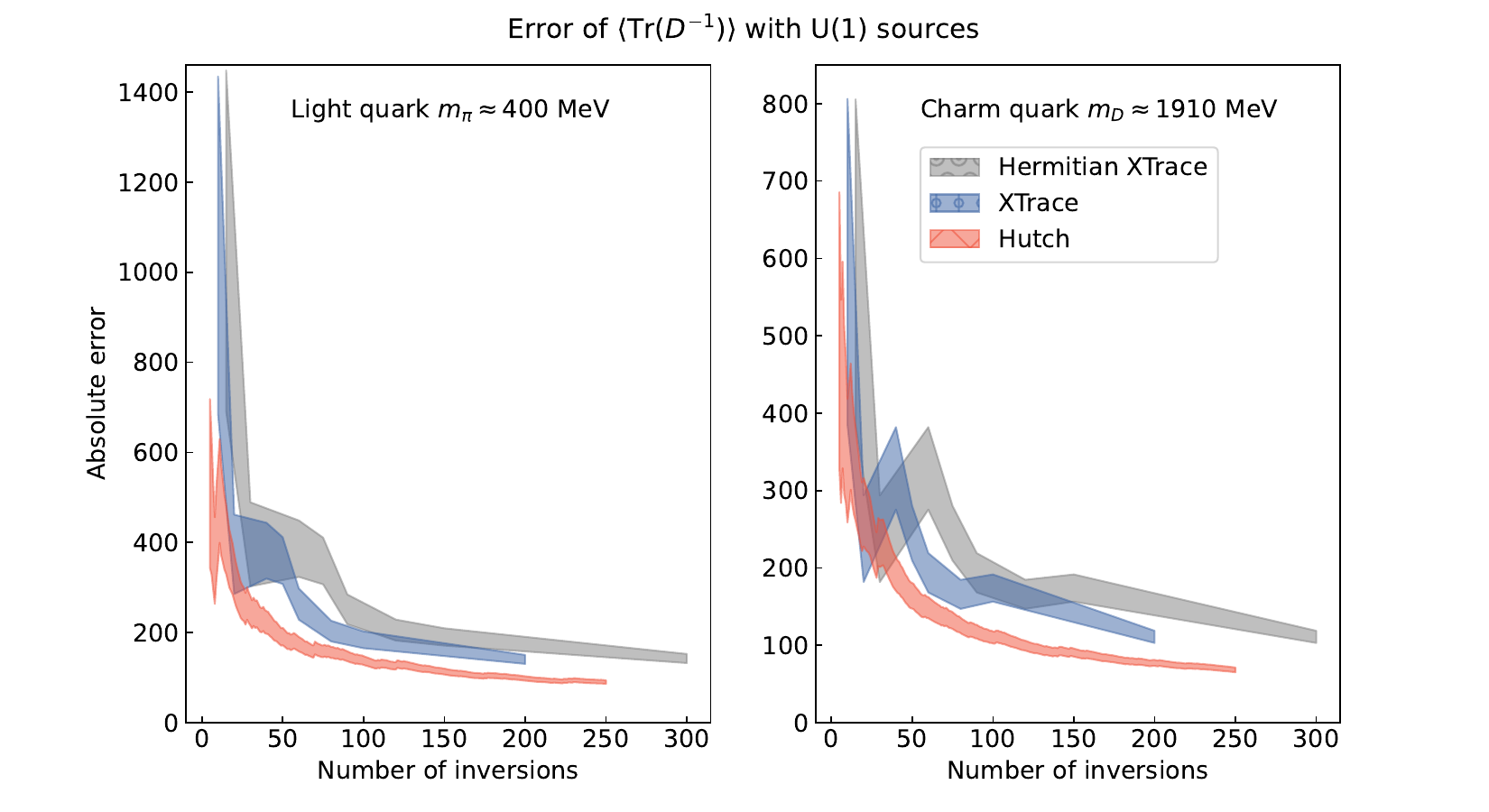}
	\caption{\label{fig:XTrace_plot}Scaling of the absolute error as a function of the number of inversions of $D$ for the different algorithms using U(1) random sources for both the quarks in the ensemble.}
\end{center}
\end{figure}

We compare the scaling of the XTrace with respect to the Girard-Hutchinson estimator for $O(200)$ number of inversions of the Dirac operator to see which of them performs better. The results are shown in Figure \ref{fig:XTrace_plot} and they clearly show that the method does not overcome the Girard-Hutchinson trace estimator both in the light quark and in the charm quark cases.

\newpage
\section{Conclusions \& Outlook}
In this study, we tested the Hutch++ and XTrace stochastic trace estimators versus the Girard-Hutchinson method to understand if they can be used as a substitute to speed up the trace computations in Lattice QCD simulations since they have been proven to always have better scaling of the error in the asymptotic regime. We did this test using a single configuration from a QCD ensemble generated by the RC$^{\star}$ collaboration. We tested both methods with a variant which we called hermitian that generates the deflation space with the hermitian part of the inverse of the Dirac operator.
 
Unfortunately, the two methods did not show any improvement in the regime of $O(200)$ inversions of the Dirac operator. We investigated the impact of the deflated term on the total trace for different pion masses in a quenched set-up but this showed that the term is still irrelevant even going to a mass close to the physical one.

 \acknowledgments 
We acknowledge Stefan Schaefer for pointing us to the Hutch++ paper and all the members of the RC$^{\star}$ collaboration for the fruitful discussions and feedback on this work.
AC's research is funded by the Deutsche Forschungsgemeinschaft (DFG, German Research Foundation) - Projektnummer 417533893/GRK2575 "Rethinking Quantum Field Theory".
The authors gratefully acknowledge the computing time granted by the Resource Allocation Board and provided on the supercomputer Lise and Emmy at NHR@ZIB and NHR@Göttingen as part of the NHR infrastructure. The calculations for this research were partly conducted with computing resources under the projects bep00085 and bep00102.

\bibliographystyle{JHEP}
\bibliography{inspire}

\end{document}